\documentclass[aps,prd,twocolumn,showpacs,showkeys,preprintnumbers]{revtex4-1}

\usepackage{graphicx}
\usepackage{relsize}
\def\babar{\mbox{\slshape B\kern-0.1em{\smaller A}\kern-0.1em
    B\kern-0.1em{\smaller A\kern-0.2em R}}}

\begin{document}


\preprint{RWTH Aachen TTK--10--04}

\title{The Mass of the Higgs Boson in the Standard Electroweak Model}

\author{Jens Erler}
\email[]{erler@fisica.unam.mx}
\affiliation{Departamento de F\'isica Te\'orica, Instituto de F\'isica, \\
              Universidad Nacional Aut\'onoma de M\'exico, 04510 M\'exico D.F., M\'exico}
\altaffiliation{Present address: Institut f\"ur Theoretische Physik E, RWTH Aachen, 52056 Aachen, Germany}

\date{\today}

\begin{abstract}
An updated global analysis within the Standard Model (SM) of all relevant electroweak precision and Higgs boson search data is presented with special emphasis on the implications for the Higgs boson mass, $M_H$.
Included are, in particular, the most recent results on the top quark and $W$ boson masses, 
updated and significantly shifted constraints on the strong coupling constant, $\alpha_s$, from $\tau$ decays
and other low energy measurements such as from atomic parity violation and neutrino deep inelastic scattering.
The latest results from searches for Higgs production and decay at the Tevatron are incorporated together with the older
constraints from LEP~2. I find a trimodal probability distribution for $M_H$ with a fairly narrow preferred 90\% CL window,
$115\mbox{ GeV} \leq M_H \leq 148\mbox{ GeV}$.
\end{abstract}


\maketitle

\section{\label{intro}Introduction}
One of the prime missions of the Large Hadron Collider (LHC) at CERN is the search for the Higgs boson.
Within the SM, its existence is solidly predicted but only some semi-quantitative theoretical constraints exist for its mass.
If the SM is the correct low-energy theory only up to a new physics scale which is itself not much larger than $M_H$,
one would find at one-loop order and neglecting all couplings other than the Higgs self-coupling $\lambda$, the "triviality" condition,
$$ {M_H^2\over v^2} \ln {M_H^2\over v^2} < {8\pi^2\over 3}, \quad\quad\quad\quad  M_H < 816 \mbox{ GeV}, $$
where $v = 246.22$~GeV is the Higgs vacuum expectation value.
This is very close to the bound obtained from the requirement of unitarity 
of the partial S-wave amplitude of elastic Goldstone boson scattering~\cite{Luscher:1988uq}, 
$$ {M_H^2\over v^2} < {16\pi\over 5}, \quad\quad\quad\quad\quad\quad\quad  M_H < 781 \mbox{ GeV}. $$
Requiring the absence of a Landau pole in $\lambda$ up to the reduced Planck scale,
$\kappa_P = 2.4 \times 10^{18}$~GeV, yields in the same approximation as above,
$$ {M_H^2\over v^2}  < {4\pi^2\over 3} \ln^{-1} {\kappa_P\over v}, \quad\quad\quad  M_H < 147 \mbox{ GeV}, $$
while a refined analysis including top quark mass, $m_t$, 
and two-loop effects gives $M_H \lesssim {\cal O}(m_t)$~\cite{Hambye:1996wb}.
Vacuum stability, {\em i.e.}, the requirement that the scalar potential is bounded below, implies the lower bound 
(driven by the large top quark Yukawa coupling), $M_H \gtrsim 130 \mbox{ GeV}$~\cite{Casas:1994qy}, 
but the vacuum would still be sufficiently long-lived (metastable) for $M_H \gtrsim 115 \mbox{ GeV}$~\cite{Isidori:2001bm}.

Quantum loop corrections affecting the multitude of electroweak precision observables 
--- most importantly the $W$ boson mass, $M_W$, and effective weak mixing angles, $\theta_W^{\rm eff}$ ---
allow to simultaneously test the model and over-constrain its free parameters including $M_H$.
Moreover, by comparing various cross-section measurements with the SM prediction as a function of $M_H$ 
one may identify preferred and disfavored regions.
In this way, CDF and D\O\ at the Tevatron~\cite{Aaltonen:2010yv} concluded that the window, 
$162\mbox{ GeV} < M_H < 166\mbox{ GeV}$, is incompatible with their data at the 95\% CL.
Likewise, the LEP~2 Collaborations~\cite{Barate:2003sz} found the 95\% CL lower limit, $M_H \geq 114.4$~GeV.

In this communication, I update the global analysis of Ref.~\cite{Erler:2000cr} where the statistical method is described in detail.
I incorporate all direct (search) and indirect (precision) data, including new radiative corrections and significant improvements 
and changes in several precision observables. 
For alternative analyses, see Refs.~\cite{Degrassi:2001tg,Flacher:2008zq},
where the latter differs mostly by the neglect of the low energy data and the treatment of LEP~2 (see Section~\ref{direct}).

\section{Experimental data}
\subsection{\label{indirect}Electroweak precision observables}
The strongest $M_H$ constraints come from asymmetry measurements determining $\sin^2\theta_W^{\rm eff}$ for leptons~\cite{LEPEWWG:2005ema} at LEP~1 and the SLC~\cite{Abe:1996ef,Abe:2000hk}, 
and from $M_W$ at LEP~2 and the Tevatron~\cite{LEPEWWG:2009jr}.
The two most precise determinations of $\sin^2\theta_W^{\rm eff}$ deviate by about 3~$\sigma$ from each other, 
but since both are statistics dominated we consider this in the following as a fluctuation.
These constraints are strongly correlated (39\%) with $m_t$, 
giving great importance to the direct determination of the latter at the Tevatron~\cite{TEVEWWG:2009ec},
$$m_t = 173.1 \pm 0.6\ ({\rm stat.}) \pm 1.1\ ({\rm syst.}) \pm 0.5\ ({\rm QCD})\ {\rm  GeV}. $$
This is converted to the $\overline{\rm MS}$-mass definition using the three-loop formula~\cite{Melnikov:2000qh}
which gives rise to the QCD error (the size of the three-loop term). 
It is assumed that this accounts for the error from interpreting the mass extracted at the Tevatron as the pole mass. 

Other $Z$ pole constraints are the $Z$ width, $\Gamma_Z$, the total hadronic peak cross section, $\sigma_{\rm had}$, 
and a set of branching ratios, $R_i$~\cite{LEPEWWG:2005ema}.
Finally, there is a wide range of low energy experiments from atomic parity violation (APV) to neutrino and polarized electron scattering.
See Ref.~\cite{Erler:2009jh} for tables of inputs almost identically to those used here.

All experimental and theoretical uncertainties and correlations are included in the fits.
The error from unknown higher order electroweak corrections is implemented 
by allowing the so-called oblique $T$-parameter~\cite{Peskin:1990zt} to float subject to the constraint $T = 0 \pm 0.02$.
Errors from different sources have been added in quadrature and in most (but not all) cases been treated as Gaussian. 
The resulting constraints are depicted in Figure~\ref{mhmt}.
Some comments on those inputs which have shifted recently or which have been discussed controversially are in order:

\begin{figure}[t]
 \includegraphics[width=250pt] {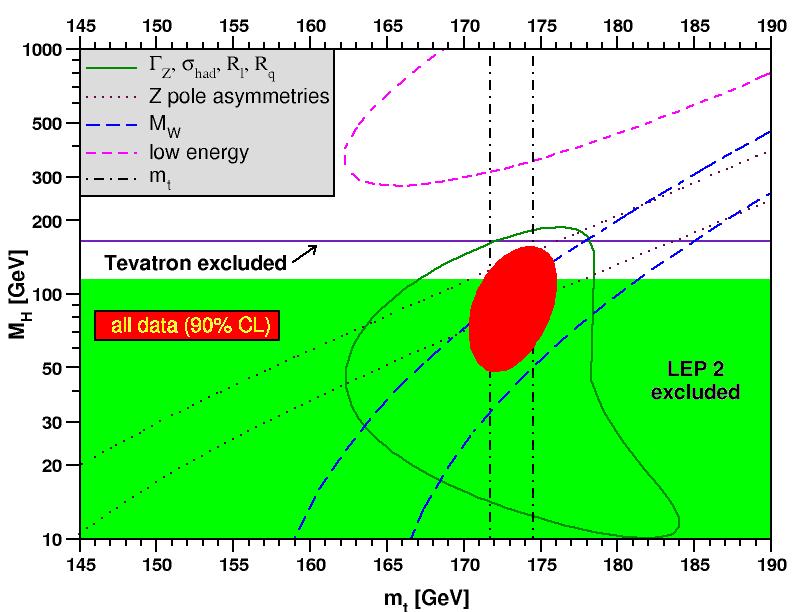}
 \caption{\label{mhmt} 1~$\sigma$ (39.35\% CL) contour lines for $M_H$ as a function of  $m_t$ for various inputs and
the solid (red) 90\% CL ellipse ($\Delta \chi^2 = 4.605$) allowed by all data. 
$\alpha_s(M_Z) = 0.1183$ is assumed except for fits including the $Z$~lineshape or low energy data. 
The lower limit from LEP~2 and the excluded window from the Tevatron (both at the 95\% CL) are also shown.}
 \end{figure}

The theoretical predictions for $M_W$ and $\sin^2\theta_W^{\rm eff}$ need 
the renormalization group evolution of the electromagnetic coupling from the Thomson limit to the weak scale.
Entering the implementation (in the FORTRAN package GAPP~\cite{Erler:1999ug}) is the $\overline{\rm MS}$ definition, 
$\hat\alpha(M_Z)$, which is updated from Ref.~\cite{Erler:1998sy} with its central value moved upwards and its uncertainty almost halved. 
The corresponding hadronic vacuum polarization effects can be translated from cross-section data for $e^+ e^- \to$ hadrons,
which in turn can be obtained by standard $e^+ e^-$ annihilation or by the high statistics (but systematics dominated) 
method~\cite{Arbuzov:1998te} of using radiative returns from a $1S$ resonance.
In addition, there are measurements of $\tau$ decay spectral functions which can be included with the appropriate isospin 
corrections~\cite{Davier:2009ag}. 
However, the results reveal some discrepancies.
The $\tau$ data imply lower values for the extracted $M_H$ of about 6\% compared to the $e^+ e^-$ data. 
This conflict is smaller than in the past and some of it appears to be experimental. 
The dominant $e^+e^- \to \pi^+\pi^-$ cross-section has been measured by CMD-2~\cite{Akhmetshin:2003zn} 
and SND~\cite{Achasov:2006vp} and the results are in good agreement with each other,
but are lower than those obtained from $\Upsilon (4S)$ radiative returns by \babar~\cite{Aubert:2004kj}. 
In turn, the latter agrees quite well with the $\tau$ analysis including the energy dependence (shape). 
In contrast, the shape and smaller overall cross-section from  $\pi^+\pi^-$ pairs radiatively returned from the $\Phi$ and detected
by KLOE~\cite{Ambrosino:2008en} differ significantly from \babar\ (a recent review on the $e^+e^-$ data is Ref.~\cite{Davier:2009zi}). 
All measurements including older data are accounted for on the basis of results from
Refs.~\cite{Davier:1998si,Davier:2009ag,Davier:2009zi}. 
The correlation with the $\mu^\pm$ magnetic moment and the non-linear $\alpha_s$ dependence of $\hat\alpha (M_Z)$ are
addressed. 
The correlation of $\hat\alpha (M_Z)$ with $\alpha_s$ has been treated by using as input (fit constraint) instead of 
$\Delta\alpha_{\rm had}^{(5)}(M_Z)$ the analogous low-energy contribution by the three light quarks, 
$\Delta\alpha_{\rm had}^{(3)}(1.8~{\rm GeV}) = (57.29 \pm 0.90) \times 10^{-4}$, and by calculating 
the perturbative and heavy quark contributions to $\hat\alpha (M_Z)$ in each call of the fits according to Ref.~\cite{Erler:1998sy}.
The error is from $e^+ e^-$ data below 1.8~GeV and $\tau$ decay data, 
from uncertainties in the isospin breaking effects (affecting the interpretation of the $\tau$ data),
from unknown higher order perturbative and non-perturbative QCD effecs; 
and from the charm and bottom quark masses.

There is extra information on $\sin^2\theta_W$ and $M_H$ in the $Z$ boson vector couplings, 
which is used best if $\alpha_s$ is constrained independently. 
For this I use the extraction of $\alpha_s$ from the $\tau$ lifetime, $\tau_\tau$, because
(i) the $\tau$ scale is low, so that upon extrapolation to the $Z$~scale, the $\alpha_s$ error shrinks by an order of magnitude; 
(ii) this scale is still high enough that the operator product expansion (OPE) can be applied; 
(iii) $\tau_\tau$ is fully inclusive and thus free of hadronization effects; 
(iv) OPE breaking effects occur only where they are kinematically suppressed;
(v) non-perturbative effects can be constrained by experimental data;
(vi) the complete four-loop order (massless) QCD expression is known; and
(vii) large effects associated with the QCD $\beta$-function can be re-summed~\cite{LeDiberder:1992te} 
in contour improved perturbation theory (CIPT).
However, while CIPT shows faster convergence in the lower calculated orders, 
doubts have been cast on the method by the observation that at least in a specific model~\cite{Beneke:2008ad} 
including theoretical constraints on the large-order behavior,
ordinary fixed order perturbation theory (FOPT) may nevertheless give a better approximation. 
The largest uncertainty arises from the truncation of the FOPT series and is taken as the $\alpha_s^4$ term. 
I find $\alpha_s(M_Z)~[\tau] = 0.1174^{+0.0018}_{-0.0016}$ which updates Ref.~\cite{Erler:2002bu}. 
The effects of using FOPT instead of CIPT, 
of using the theoretically better motivated spectral functions of Ref.~\cite{Maltman:2008nf} in place of previous results, 
and of including the four-loop result~\cite{Baikov:2008jh}, all significantly reduce the extraced $\alpha_s$ value. 

 \begin{figure}[t]
 \includegraphics[width=250pt] {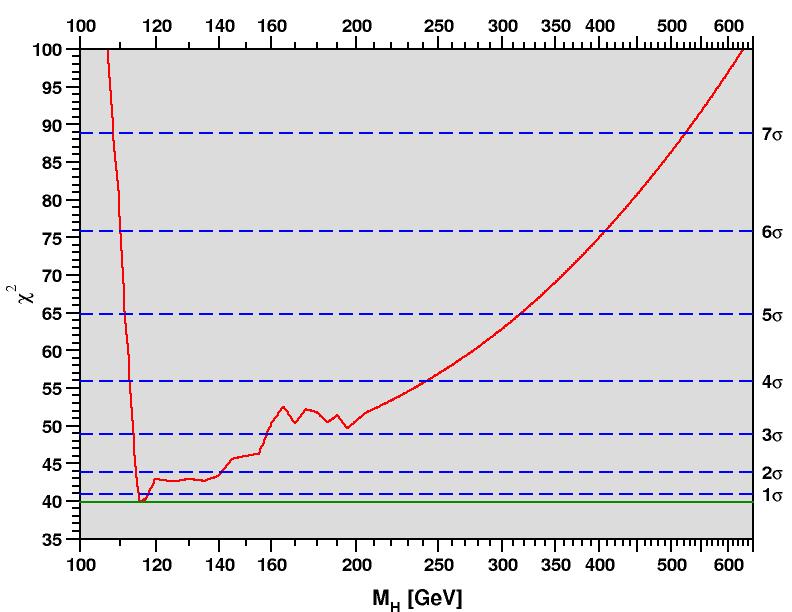}
 \caption{\label{chi2plot} $\chi^2$ distribution of $M_H$ from all data. 
The horizontal solid (green) line marks the $\chi^2$ minimum,  $\chi^2_{\rm min} = 39.88$, 
while the dashed (blue) lines refer to integer values of $\sqrt{\chi^2 - \chi^2_{\rm min}}$.
The combined effects of precision, LEP~2, and Tevatron data result in the pronounced dip at $M_H = 115.8$~GeV.
Values of $M_H > 141$~(158)~GeV are excluded at the 2 (3) $\sigma$ level.}
 \end{figure}

There are precise APV experiments in Cs~\cite{Wood:1997zq,Guena:2004sq} and Tl~\cite{PhysRevLett.74.2654,Vetter:1995vf}, 
where the error associated with atomic wave functions is quite small for Cs~\cite{Ginges:2003qt}. 
The extracted weak mixing angle (in the $\overline{\rm MS}$-scheme), $\hat{s}^2_W = 0.2314 \pm 0.0014$,
now agrees perfectly with $\hat{s}^2_W = 0.23116 \pm 0.00013$ from the SM fit,
where the theoretical effects in Refs.~\cite{Ginges:2003qt,Porsev:2009pr} 
together with an update of the SM calculation~\cite{Erler:2003yk} removed an earlier 2.3~$\sigma$ deviation from the SM.

Neutrino-nucleus deep inelastic scattering ($\nu$-DIS) 
is dominated by the NuTeV result~\cite{Zeller:2001hh} for the on-shell weak mixing angle, $s^2_W = 0.2277\pm 0.0016$, 
which initially was 3.0~$\sigma$ higher than the SM prediction, $s^2_W = 0.22292 \pm 0.00028$.
Since then a number of experimental and theoretical developments shifted the extracted $s^2_W$,
most of them reducing the discrepancy:
(i) NuTeV also measured~\cite{Mason:2007zz} the difference between the strange and antistrange quark momentum distributions, 
$S^- = 0.00196 \pm 0.00143$.
The effect of $S^- \neq 0$ on the NuTeV value for $s^2_W$ has been studied in Ref.~\cite{Zeller:2002du}, 
and the $S^-$ above shifts $s^2_W$ by $-0.0014 \pm 0.0010$. 
In view of theoretical arguments favoring a much smaller or negligible effect, 
I take half of the above shift as an estimate of both the $S^-$ effect and the associated error.
(ii) The measured branching ratio for $K_{e3}$ decays enters crucially 
in the determination of the $\nu_e (\bar\nu_e )$ contamination of the $\nu_\mu (\bar\nu_\mu)$ beam.
Since the time of Ref.~\cite{Zeller:2001hh} it has changed by more than 4~$\sigma$, so that a move of $s^2_W$ by $+0.0016$ 
is applied and the corresponding $\nu_e (\bar\nu_e )$ error decreased by a factor of $2/3$.
(iii) Parton distribution functions (PDFs) seem to violate isospin symmetry at levels much stronger than expected.
While isospin symmetry violating PDFs are currently not well constrained phenomenologically,
the leading contribution from quark mass differences turns out to be largely model-independent~\cite{Londergan:2003ij}
and a shift, $\delta s^2_W = - 0.0015 \pm 0.0003$~\cite{Martin:2004dh,Bentz:2009yy}, is applied.
(iv) QED splitting effects also violate isospin symmetry, shifting $s^2_W$ by $- 0.0011$~\cite{Gluck:2005xh} 
with a 100\% assigned error (the sign is model-independent).
(v) The isovector EMC effect~\cite{Cloet:2009qs} model-independently reduces the discrepancy,
shifting $s^2_W$ by $- 0.0019 \pm 0.0006$~\cite{Bentz:2009yy}.
(vi) The extracted $s^2_W$ may also shift significantly when analyzed using the most recent QCD~\cite{Dobrescu:2003ta},
QED and electroweak~\cite{Diener:2005me} radiative corrections,
but their precise impact will be revealed only after the NuTeV data have been re-analyzed with a new set of PDFs.
Remaining radiative corrections have been estimated~\cite{Diener:2005me} to induce an error of $\pm 0.0005$ in $s^2_W$.
With these corrections, the global $\nu$-DIS average in terms of the weak mixing angle is $s^2_W = 0.2254 \pm 0.0018$.

 \begin{figure}[t]
 \includegraphics[width=250pt] {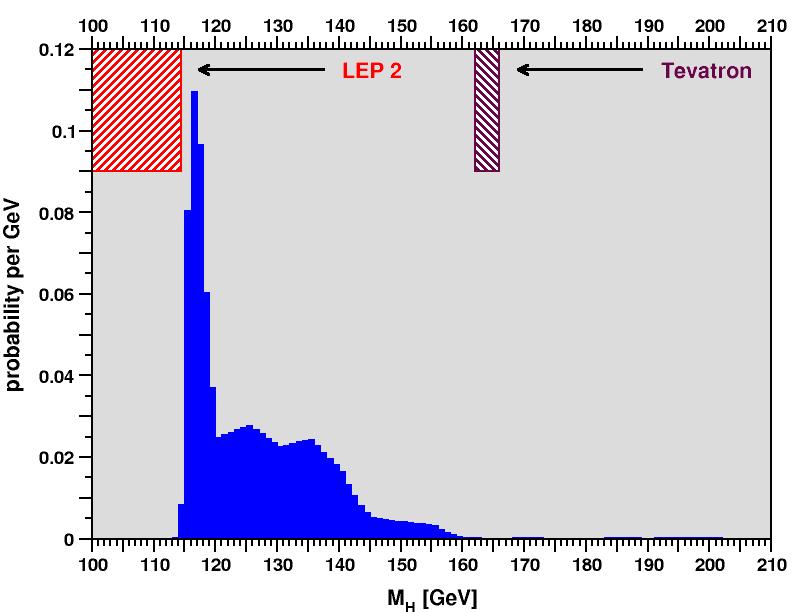}
 \caption{\label{mhplot} Probability distribution of $M_H$ subject to all data. 
 The nominal 95\% CL exclusion ranges~\cite{Barate:2003sz,Aaltonen:2010yv} from LEP~2 and the Tevatron are also indicated.}
 \end{figure}

\subsection{\label{direct}Collider searches}
At LEP~2 with energies up to $\sqrt{s} \approx 209$~GeV, the Higgs boson was searched for 
in the dominant ($\approx 74\%$) $b\bar{b}$ decay channel, produced in the Higgsstrahlung process, $e^+ e^- \to ZH$. 
In addition, the $H \to \tau^+ \tau^-$ channel ($\approx 7\%$) was studied for the $Z$ boson decaying into two jets. 
The combination~\cite{Barate:2003sz} of the four experiments, all channels and all $\sqrt{s}$ values, 
resulted in the nominal lower bound, $M_H \geq 114.4$~GeV.  
However, the combined data are neither particularly compatible with the hypothesis $M_H = 115$~GeV (15\% CL),
nor with background only (9\% CL).
The reason is that the results by ALEPH are by themselves in very good agreement with $M_H \approx 114$~GeV 
(due to an excess in the 4-jet channel) thereby strongly rejecting the background only hypothesis,
while the results based on the other channels and experiments (especially DELPHI) are incompatible with any signal.
Overall, a signal for $115\mbox{ GeV} \leq M_H \leq 119.5\mbox{ GeV}$ is favored by the data, but not with high significance. 

\begin{table*}
\caption{\label{CLs} Upper and lower bounds for $M_H$ [in GeV] at various confidence levels (CLs).
Shown are the Bayesian limits defined as the central intervals of the integrated probability density function (see Figure~\ref{mhplot}).
Also show (in parentheses and for comparison only) are the limits derived from simple frequentist (maximum likelihood) reasoning
(see also Figure~\ref{chi2plot}). 
The two values in the first line are for the median and the best fit, respectively. 
They differ because the distribution is not symmetric.}
\begin{ruledtabular}
\begin{tabular}{llllllr} 
CL [\%]      & $1 - $CL [\%] & lower bound & (frequentist) & upper bound & (frequentist) & interpretation \\ 
\hline
50              & 50                 & 123.6            & (115.8)         & 123.6            & (115.8)         & median ($\chi^2_{\rm min}$) \\
84.134       & 15.866          & 116.1            & (114.9)         & 137.6            & (117.7)         & 1$\sigma$ range \\
95              & 5                   & 115.1            &                     & 148               &                     & 90\% CL range \\
97.725       & 2.275            & 114.8            & (114.3)         & 155               & (141)            & 2$\sigma$ range \\
99.5           & 0.5                & 114.3            &                     & 197               &                     & 99\% CL range \\
99.865       & 0.135            & 113.9            & (113.5)         & 217               & (158)            & 3$\sigma$ range \\
99.95         & 0.05              & 113.6            &                     & 232               &                     & 99.9\% CL  range \\
99.9968     & 0.0032          & 112.9            & (112.6)         & 276               & (240)            & 4$\sigma$ range \\
99.9995     & 0.0005          & 112.4            &                     & 304               &                     & 99.999\% CL range \\
99.999971 & 0.000029      & 111.5            & (111.3)         & 349               & (315)            & 5$\sigma$ range \\
\end{tabular}
\end{ruledtabular}
\end{table*}

The LEP~2 results can be included by adding the solid line for the observed log-likelihood ratio (LLR$_{\rm obs}$) shown 
in Figure~1 of Ref.~\cite{Barate:2003sz} to the $\chi^2$-function derived from the precision observables in Section~\ref{indirect}.
The quantity LLR$_{\rm obs}$ is defined as $-2 \ln Q(M_H)$, where $Q(M_H)$ is 
the ratio of the likelihood for the signal of a particular $M_H$ hypothesis plus the background to that of the background alone. 
This treatment is rigorous in the limit of large data samples and serves as a good approximation otherwise. 
It is emphasized that treating the LEP~2 results as a step function with threshold at the nominal lower $M_H$ bound
is a poor approximation whenever there is a noticeable upward fluctuation in the data beyond that threshold
and results in systematic and significant upwards shifts of the upper bounds (compare, {\em e.g.}, with Ref.~\cite{Flacher:2008zq}).

At the Tevatron running at  $\sqrt{s} = 1.96$~TeV, the Higgs boson can conceivably be produced in association with $W$ or $Z$ bosons,
$p\bar{p} \to W/Z H$ (the counterpart of Higgsstrahlung at LEP~2), 
or through gluon ($gg \to H$) or vector boson ($q\bar{q} \to q'\bar{q}' H$) fusion.
The studied decay channels besides $H \to b\bar{b}$ and $H \to \tau^+ \tau^-$ are the one-loop decay $H \to \gamma\gamma$ as well as
(dominant for $M_H \gtrsim 140$~GeV) $H\to W^+ W^-$.
For their combination~\cite{TEVNPH:2009je} CDF and D\O\ analyzed 90 individual processes.
As was the case at LEP~2, but somewhat more significantly, the low mass Higgs region is favored by the data.
This is pronounced around 115~GeV, but persists until $M_H = 155$~GeV.
On the other hand, the range $155\mbox{ GeV} \leq M_H \leq 197\mbox{ GeV}$ is disfavored especially in the nominal exclusion window.
The Tevatron results are incorporated by adding the LLR$_{\rm obs}$ column given in Table XIX of Ref.~\cite{TEVNPH:2009je}
as another contribution to $\chi^2$.

\section{Results}

The following results are based on a Bayesian treatment~\cite{Erler:2000cr} which is particularly adequate for parameter estimation
(as opposed to hypothesis testing).  
The notorious objection of the necessity of a prior distribution notwithstanding, 
Bayesian data analysis provides a first principles setup with strong emphasis on the entire posterior density~\cite{citeulike:105949},
It is given by,
\begin{equation}
  p(M_H) = e^{-\chi^2_{\rm indirect}/2}\ Q_{\rm LEP~2}\ Q_{\rm Tevatron}\ M_H^{-1},
\end{equation}
where the first factor is from the precision data, while the factors of $Q = Q(M_H)$ are as described in Section~\ref{direct}.
The last factor is the (improper) non-informative prior density chosen such that the variable $\ln M_H$ has a flat prior 
which one can argue is the most conservative (least informative) for a variable defined over the real numbers.
Alternatively choosing a flat prior in $M_H$ itself ({\it i.e.,\/} dropping the factor $M_H^{-1}$) increases, for example, 
the 95\% CL upper limit by modest 3 GeV because small $M_H$ values are penalized {\it a priori}.
The 90\% preferred range (95\% CL lower and upper bounds) for $M_H$ is given by,
\begin{equation}
  115\mbox{ GeV} \leq M_H \leq 148\mbox{ GeV},
\end{equation}
with corresponding bounds for other CLs shown in Table~\ref{CLs}.
The Table also shows that a Higgs boson discovery with $M_H \geq 350$~GeV would simultaneously mean 
the indirect discovery of new weak scale physics which would have to bridge the gap between the physical Higgs mass 
and the mass derived from the current data when assuming the validity of the SM.

Figure~\ref{chi2plot} shows the distribution of $\chi^2 \equiv - 2 \ln p(M_H)$, while
Figure~\ref{mhplot} represents $p(M_H)$ binned in 1~GeV steps.  
Only by virtue of $\chi^2_{\rm indirect}$ is it possible to obtain a proper $p(M_H)$ density. 
A trimodal distribution emerges with a tall peak since the searches of both, 
LEP~2 and the Tevatron, see some excess events hinting at $M_H < 120$~GeV, 
and also since the high energy precision data favor $M_H = 87^{+35}_{-26}$~GeV~\cite{LEPEWWG:2009jr}.
This $M_H$ value agrees well with Ref.~\cite{Flacher:2008zq} ($M_H = 80^{+30}_{-23}$~GeV)
and the global fit here ($M_H = 91^{+31}_{-24}$~GeV).
The other two modes are due to the precision data in regions where $Q_{\rm Tevatron}$ is basically flat.

\begin{acknowledgments}
Work supported by CONACyT project 82291--F and by the German Academic Exchange Service (DAAD).
It is a pleasure to greatly acknowledge the hospitality and support extended by the Institute for Theoretical Physics E 
of the RWTH Aachen and to thank Michael Kr\"amer and Peter Zerwas for many helpful discussions.
\end{acknowledgments}

\bibliography{HiggsPDF}

\end{document}